\documentclass{PoS}

\title{Warped Views on the Large Hadron Collider}

\ShortTitle{Warped Views on the LHC}

\author{\speaker{Jos\'e Santiago}\thanks{
Work supported by projects
FPA2006-05294, FQM 101, FQM 03048 and by MICINN through a Ram\'on y
Cajal contract.}\\
        CAFPE and Departamento de F\'{\i}sica Te\'orica y del
        Cosmos,\\
        Universidad de Granada, E-18071 Granada, Spain\\
        E-mail: \email{jsantiago@ugr.es}}

\abstract{Models with warped extra dimensions, and their strongly
  coupled duals, offer a nice solution to the hierarchy problem and a
  very appealing realisation of flavour. Compatibility with the
  very stringent electroweak and flavour tests have made a
  generic picture emerge, with a composite Higgs, partial
  compositeness and custodial symmetry as the main ingredients. We
  review the main features of this picture and discuss
  how -and when- models with warped extra dimensions could be
  discovered at the Large Hadron Collider.}

\dedicated{I dedicate this IUPAP C11 Young Scientist Prize (theory)
 to my family and especially to Blanca and Lucas}

\FullConference{35th International Conference of High Energy Physics -
  ICHEP2010,\\ 
		July 22-28, 2010\\
		Paris France}

\begin{document}

\section{Introduction}

Models with warped extra dimensions~\cite{Randall:1999ee} are a very
interesting candidate for new physics at the TeV scale. They can stabilize
the electroweak (EW) scale by means of gravitational red-shift and
even shed light 
on other mysteries of our current understanding of particle physics,
including the structure of flavor
(see~\cite{Grossman:1999ra,Gherghetta:2000qt} 
for early references, more recent work can be found
in~\cite{neubert:ichep}) or dark
matter~\cite{Agashe:2004ci}, for instance. 
Furthermore, they provide weakly coupled duals to strongly coupled 4D
models of electroweak symmetry breaking
(EWSB)~\cite{ArkaniHamed:2000ds}. Indeed, even if one could 
argue that strong EWSB is much more general than the models which can
be parameterized through warped extra dimensions, the ability to
perform explicit calculation in the latter makes them an ideal
playground to understand generic features of models of strong EWSB.

\section{The New Standard Model from Warped Extra Dimensions} 

Models with warped extra dimensions, being dual to strongly coupled
models, are likely to be very  constrained by electroweak and flavor
precision observables. In fact, even though the main model building
ingredients were developed in some of the earliest
works~\cite{Gherghetta:2000qt}, it took several years to find a
canonical model with all the required features. The model in question
is a slice of AdS$_5$ space with metric
\begin{equation}
ds^2 = 
a^2(z) \big[ \eta_{\mu\nu} \mathrm{d}x^\mu \mathrm{d}x^\nu - dz^2
  \big],
\end{equation}
where the warp factor reads $a(z)=L_0/z$. 
The extra dimension is bounded by two three-branes, located at $z=L_0$
and $z=L_1$, and called UV brane and IR brane, respectively ($L_0 <
L_1$). This metric is a solution of Einstein's equations provided
the 5D cosmological constant and the brane tensions satisfy
$\Lambda=-V_{UV}/L_0 = V_{IR}/L_0=-24 M^2L_0^{-2}$,
where $M$ is the 5D Planck mass~\cite{Randall:1999ee}. 
The hierarchy problem is solved if all
the 5D parameters (except for $L_1$) 
are of the order of the 4D Planck mass 
$\sim M_{\mathrm{Pl}}$, 
the Higgs is localized near the IR brane and we have the hierarchy
$L_0/L_1 \approx \mathrm{TeV}/M_{\mathrm{Pl}}$. This hierarchy can in
principle be generated with a mild fine-tuning by means of a bulk
scalar~\cite{Goldberger:1999uk}. Once the Higgs is localized close to
the IR brane, fermion masses will be naturally hierarchical if they
are allowed to propagate in the bulk of the extra dimension, with
lighter Standard Model (SM) 
fermions being localized close to the UV brane and heavier (top and
bottom) fermions being closer to the IR brane. This construction
automatically produces hierarchical mixing angles and, thanks to the
warping, an extra flavor protection that ensures flavor violation to
scale with fermion masses or mixing
angles~\cite{Grossman:1999ra,neubert:ichep}. This scaling is typically
enough to keep flavor observables under control (although some
structure might be needed for full compatibility,
see~\cite{neubert:ichep}). The reason for this is that the warping
forces the lightest Kaluza-Klein (KK) excitations of bulk fields (and
therefore the relevant new physics) to be localized close to the IR
brane and therefore new physics effects are naturally sizable for
heavy SM fields (which live close to the Higgs, thus the IR brane) and
suppressed for light SM particles. 

Tree level corrections to electroweak observables from the heavy
Kaluza-Klein modes appear in the form of corrections to the SM gauge
boson self-energies, to SM fermion gauge couplings and in the
form of four-fermion interactions.
If we assume the light fermions to be localized at the 
UV brane (which thanks to the warping is an excellent approximation), 
all these corrections turn out to be universal and can be encoded by
means of field redefinitions in terms of the four oblique
parameters of Ref.~\cite{Barbieri:2004qk}, $\hat{T},\hat{S},W,Y$. 
The final result (see~\cite{Davoudiasl:2009cd} 
for a recent discussion of constraints
on models with warped extra dimensions and their phenomenology) 
is that $\hat{T}$ is enhanced by the ``volume'' of the extra dimension 
$\log L_1/L_0
\approx \log M_{\mathrm{Pl}}/\mathrm{TeV} \approx 35 $, 
whereas $W,Y$ are volume suppressed. Assuming for instance a boundary
Higgs, we have~\cite{Davoudiasl:2009cd}
\begin{equation}
\hat{T}_{\mathrm{bound.}}
=\frac{g^{\prime\, 2}}{g^2} \frac{m_W^2 L_1^2}{2} \log
\frac{L_1}{L_0},
\quad \hat{S}_{\mathrm{bound.}}= \frac{m_W^2 L_1^2}{2}, \quad
W_{\mathrm{bound.}}= Y_{\mathrm{bound.}} = 
\frac{m_W^2L_1^2}{4 \log \frac{L_1}{L_0}},
\end{equation}
where $g,g^\prime$ are the $SU(2)_L \times U(1)_Y$ couplings and $m_W$
is the $W$ mass. If the Higgs is allowed to propagate in the bulk,
$\hat{T}$ and $\hat{S}$ get a bit smaller whereas $W$ and $Y$ remain
the same. For instance, using a quadratic profile for the Higgs we
obtain
\begin{equation}
\hat{T}_{\mathrm{Bulk}}=\frac{1}{3}
\hat{T}_{\mathrm{bound.}},
\quad
\hat{S}_{\mathrm{Bulk}}=\frac{3}{4}
\hat{S}_{\mathrm{bound.}}.
\end{equation}
The only relevant exception to the universality of the new physics in these
models comes with the bottom quark. Due to the heaviness of the
top the left-handed doublet cannot be localized very close to the UV
brane and therefore, $b_L$ couplings receive in general sizable
corrections. In fact the correction to the $Z b_L \bar{b}_L$
coupling is also volume enhanced
(see~\cite{Davoudiasl:2009cd}). Fortunately, both volume enhanced
contributions can be canceled by means of a custodial
symmetry~\cite{Agashe:2003zs} 
(see~\cite{Carena:2003fx,Cabrer:2010si} 
for possible alternatives), which consists of
enlarging the SM gauge symmetry to 
\begin{equation}
SU(2)_L \times U(1)_Y \to 
SU(2)_L \times SU(2)_R \times U(1)_X \times P_{L,R},
\end{equation}
where the last term is a discrete symmetry exchanging the left and
right sectors. Thanks to the custodial symmetry $\hat{T}$ vanishes
exactly at tree level (in the limit of UV localized light fermions)
and so does also the volume enhanced correction to the $Z b_L
\bar{b}_L$ coupling. The main constraints come then from the tree
level contribution to $\hat{S}$ and the one loop (calculable)
contributions to $\hat{T}$ and $Z b_L \bar{b}_L$. A detailed
analysis~\cite{Carena:2006bn} which included the most relevant one
loop corrections to the EW precision observables (see~\cite{Delaunay:2010dw} 
for an updated study)
showed that, assuming flavor is
explained by wave-function localization, the generic bound on new vector
bosons in these models is
\begin{equation}
M_{\mathrm{Gauge}} \gtrsim 3.5\mbox{ TeV},\label{Mgauge}
\end{equation}
while new fermions can be sensibly lighter than that, provided the
spectrum is rich enough~\cite{Anastasiou:2009rv}.

\section{Discovering Warped Extra Dimensions at the LHC}

\subsection{New Vector Resonances}

New vector resonances are the smoking gun of warped extra
dimensions/strong EWSB. However, the discussion in the previous
section shows that in our canonical model with warped extra
dimensions, the new bosonic resonances are not the typical
$Z^\prime$. They are relatively heavy and have
suppressed couplings to light fermions, thus reducing their
production cross section and their decay to leptons. 
Furthermore, their large coupling to heavy SM particles, mainly the
top and the Higgs and longitudinal components of the EW bosons, makes
them quite broad with widths that quickly approach a sizable fraction
of their mass for multi-TeV masses. To make things even more
complicated (or interesting), the large mass of the vector resonances
makes their decay products ($t,H,V_L$) 
highly boosted and therefore standard methods of top/Higgs/gauge
boson reconstruction become very inefficient due to the strong 
collimation of their decay products.

Taking into account all these properties, several groups have studied
the LHC reach for the vector resonances of models with warped extra
dimensions, including Kaluza-Klein excitations of the 
gluons~\cite{Agashe:2006hk}, 
charged and neutral EW bosons~\cite{Agashe:2007ki}
and even gauge
bosons of the coset space~\cite{Agashe:2009bb} 
characteristic of models in which the Higgs
is a pseudo-Goldstone boson~\cite{Agashe:2004rs}.
These analyses give the somewhat discouraging results collected in
table~\ref{table:MG}. More sophisticated
methods to deal with boosted objects~\cite{Rehermann:2010vq} can
result in an extra $0.5$ TeV reach in the case of KK gluons.
Also, current studies assume that the only open channels are into SM
particles. As we will discuss in the next section, new fermionic
resonances could be light enough to participate in the decays of the
gauge boson KK modes, thus altering the result of the analyses
presented here (see for instance~\cite{Carena:2007tn}).  
\begin{table}
\begin{center}
\begin{tabular}{l|c||l|c}
type & discovery limit (100 fb$^{-1}$)  &
type & discovery $\mathcal{L}$ (mass) 
\\
\hline 
Gluon & 4 TeV \cite{Agashe:2006hk} 
& top cust. & 0.16-1.9 fb$^{-1}$ (500 GeV)\cite{Contino:2008hi} 
| 100 fb$^{-1}$ (1.5
TeV)\cite{Mrazek:2009yu}  \\
$Z^\prime$ & 2 TeV \cite{Agashe:2007ki} 
& u, d cust.& 1 fb$^{-1}$, $\sqrt{s}=7$ TeV  (1 TeV)~\cite{LHC}\\
$W^\prime$ & 2 TeV \cite{Agashe:2007ki} 
& u, d cust.& 100 fb$^{-1}$, $\sqrt{s}=14$ TeV  (3-4 TeV)~\cite{LHC}\\
Coset & 2 TeV ($3\,\sigma$)  \cite{Agashe:2009bb}
& $\tau$ cust. & 300 fb$^{-1}$ (480 GeV)~\cite{delAguila:2010es} \\
\end{tabular}
\caption{Discovery limit for different types of gauge boson and
  fermion KK modes
  in models with warped extra dimensions at the LHC. Unless otherwise
  stated the results are for $\sqrt{s}=14$ TeV.} 
\label{table:MG}
\end{center}
\end{table} 
The bottom line is that discovering the
new bosonic resonances predicted by models with warped extra
dimensions at the LHC will take time and ingenuity and might require
an upgrade 
in energy or luminosity. Luckily, our canonical model with warped
extra dimensions can accommodate and in fact it typically does, new
light fermionic resonances, some of which can be much more easily (and
much earlier) discovered
at the LHC.

\subsection{New Fermionic Resonances: Fermion Custodians}

Due to the enlarged custodial symmetry, new fermionic resonances
(fermion custodians) will
come in these models in full multiplets of the custodial
group. In order to avoid new exotic massless fermions, the
corresponding 5D fields have twisted boundary conditions and can
therefore result in ultra-light
modes~\cite{DelAguila:2001pu,Agashe:2004ci}. These ultra-light modes
are typically associated to heavier SM fields, whose zero modes are
localized closer to the IR brane (in the 4D language they are
typically associated to more composite SM fermions). 
Indeed, top custodians can be
shown to naturally appear in many models of warped extra dimensions
with custodial symmetry and play a crucial role in triggering EWSB and
providing compatibility with EW precision
tests~\cite{Cacciapaglia:2006gp,Carena:2006bn,Anastasiou:2009rv}.   
Top custodians are new, relatively light vector-like quarks with
a large coupling to the top and the SM gauge bosons or the Higgs. 
The production of
these states at the LHC has been studied
in~\cite{Contino:2008hi}
 for pair
production and~\cite{Mrazek:2009yu} for single production with a
result that just $0.16-1.9$ fb$^{-1}$ of integrated luminosity (with
$\sqrt{s}=14$ TeV) should suffice to discover top custodians of mass
$M=500$ GeV in pair production and masses up to $M\approx 1.5$ TeV could be
discovered through single production with 300 fb$^{-1}$ (the details
of the model can change slightly this ultimate reach in either
direction). 

We have argued that relatively light top custodians typically appear
in models with warped extra dimensions. However, the extension of the
custodial symmetry with the discrete $P_{LR}$ symmetry, which has been used
to protect the $Z b_L \bar{b}_L$ coupling, opens the door to light
custodians for the lighter SM fermions, whose couplings can be protected
by the same symmetry~\cite{Blanke:2008zb}. The possibility of light
valence quark (u and d) custodians was considered for the first time
in~\cite{Atre:2008iu}. A specific realization consists of two degenerate
vector-like electroweak doublets with hypercharges $7/6$ and $1/6$,
respectively, with identical Yukawa couplings to the up quark and no
further couplings to the SM fermions (in the basis in which all
SM flavor mixing occurs in the charge $-1/3$ sector). Note that the
apparently fine-tuned situation of exact degeneracy and equality of
couplings can be the result of the custodial symmetry and in fact
appears in some particular realizations of models with warped extra
dimensions~\cite{Carena:2006bn}. In this model, the SM quark couplings
are not modified by operators of dimension six and the leading,
dimension eight correction is suppressed by two powers of the up quark
mass. Similarly, the first correction that violates flavor is
proportional to the up quark mass and therefore under control
(see~\cite{Atre:2008iu} for details). This implies that the new quarks
can be relatively light and have a large mixing with the up quark,
which results in a large coupling to the SM gauge (or Higgs) bosons
and the up quark.~\footnote{These couplings are mildly constrained by
  the unitarity of the mixing matrices and by one loop contributions
  to the $S$ parameter. For instance, the coupling of one of the
  charge $2/3$ new quarks to the Z and the up quark has to be
  $\lesssim 0.7 g/(2 c_W)$ with $c_W$ the cosine of the weak
  angle~\cite{LHC}.} This large coupling makes single production the best
channel for discovery. An analysis of the Tevatron reach for such new
quarks~\cite{Atre:2008iu} 
shows that up to $M\approx 700$ GeV can be discovered with
order one Yukawa couplings and 8 fb$^{-1}$ integrated luminosity. 
The study of the LHC reach is currently under way but preliminary
results show that the early run with 1 fb$^{-1}$ at 7 TeV can be
competitive or even improve on the Tevatron reach whereas masses of up
to $3-4$ TeV could be discovered with 100 fb$^{-1}$ and $\sqrt{s}=14$
TeV~\cite{LHC}.

Finally, the possibility of new fermion custodians is not restricted to
quarks. In fact, it has been recently shown~\cite{delAguila:2010vg}
that if an $A_4$
discrete symmetry  is used to generate tri-bimaximal mixing in
models with warped extra dimensions~\cite{Csaki:2008qq}, the
tau lepton might  be more localized towards the IR brane than
naively expected. The reason is that both the charged lepton Yukawa
couplings and lepton flavor violation are suppressed by the scale of
$A_4$ breaking and compatibility of the latter with observation
implies an extra suppression on the lepton Yukawas which requires a
stronger localization to reproduce the tau mass.
  This makes light tau custodians a natural occurrence
in such models. The structure is identical to the one we have
described for the up quark, with two new degenerate vector-like
leptons which are doublets of the EW symmetry with hypercharges $-1/2$
and $-3/2$, respectively. The custodial symmetry again protects the
tau couplings and we checked in~\cite{delAguila:2010vg} 
that both EW precision tests
and flavor constraints are typically under control.
The LHC reach for pair production of tau custodians has been recently
studied in~\cite{delAguila:2010es}. Each tau custodian 
decays into a SM gauge boson or the Higgs and a tau lepton. Requiring 
at least
one of the custodians to decay into a leptonic $Z$, the two taus to
decay also leptonically and the other boson to decay
hadronically results in four leptons, two jets and missing energy in
the final state. Note that, due to the large boost of the two taus,
the event can be completely reconstructed assuming full collimation of the
decay products, despite the four neutrinos in the
final state. The large number of leptons in the final state, together
with pair production of relatively heavy objects makes it
easy to reduce the background to negligible levels. However, the small
(EW) production cross section and the reduction due to the leptonic
branching fractions makes the reach, after 300 fb$^{-1}$ at
$\sqrt{s}=14$ TeV, $M\approx 480$ GeV in this channel. The results for
fermion custodian searches at the LHC are summarized in Table~\ref{table:MG}.

\section{Conclusions and Caveats}

Models with warped extra dimensions offer a nice solution to the
hierarchy problem and a very appealing explanation of the flavor
structure observed in nature. Electroweak and flavor constraints lead
us to a canonical model with custodial symmetry and 
a relatively high scale ($\sim 3.5$ TeV) of new bosonic
resonances which are not easy to find at the LHC. The reason is that,
besides being relatively heavy, they have small couplings to light SM fermions
and therefore reduced production cross sections but strong coupling to
the top and longitudinal EW gauge bosons, which makes them
broad. Furthermore, they decay into boosted objects that require
special techniques to be disentangled from the QCD background.
Custodial symmetry, a natural ingredient in these models to tame large
corrections to the $T$ parameter and the $Z b_L \bar{b}_L$ coupling,
imply the existence of new fermionic resonances, the fermion
custodians, that can be light and couple strongly to the SM particles.
Both top custodians and valence (up or down) quark custodians can be
discovered very early at the LHC (and even at the Tevatron for the
latter) if they are light enough. Tau custodians lighter than $\approx
500$ GeV could be discovered at the LHC, although a large luminosity
might be needed on the upper end of this mass range. Only after a long
high energy (14 TeV) high luminosity (100-300 fb$^{-1}$) run 
the first colored vector resonances could be discovered whereas a much
longer run would be needed for the discovery of electroweak new
resonances.

The discussion in this talk has focused on one particular scenario
with warped extra dimensions which, although arguably quite natural
could be different from the one realized in nature. For instance,
simple modifications of the background can lead to reduced electroweak
and flavor constraints~\cite{Falkowski:2008fz,Batell:2008me} 
and therefore make discovery at the
LHC much simpler in principle. Similarly, Higgsless models (see 
\cite{Csaki:2003dt} and the first reference
in~\cite{Cacciapaglia:2006gp}) 
which by giving up on the geometrical realization of flavor are
compatible at tree level with a much smaller scale of new physics,
will have a completely different phenomenology. 
Even custodial symmetry might be
expendable in some constructions~\cite{Cabrer:2010si} 
in which new light fermionic
resonances are not necessarily natural. Even in the class of models we
have discussed, we could not cover all interesting implications at the
LHC. In particular those related to new features in the Higgs and
longitudinal gauge boson sector (see for
instance~\cite{Djouadi:2007fm,Csaki:2000zn} 
and the last
reference of~\cite{Blanke:2008zb}).
The properties of all these models are worth studying in the light of
the exciting times ahead of us.

\end{document}